# Model of ionic currents through microtubule nanopores and the lumen

Holly Freedman[a,b], Vahid Rezania[a,c], Avner Priel[a,b], Eric Carpenter[a,b], Sergei Y. Noskov[d] and Jack A. Tuszynski[a,b*]

[a]Department of Oncology, University of Alberta, Cross Cancer Institute, Edmonton, AB;

[b]Department of Physics, University of Alberta, Edmonton, AB;

[c]Department of Physical Sciences, Grant MacEwan University, Edmonton, AB;

[d]Institute of Biocomplexity and Informatics, Department of Biological Sciences, University of Calgary, Calgary, AB

**Abstract**

It has been suggested that microtubules and other cytoskeletal filaments may act as electrical transmission lines. An electrical circuit model of the microtubule is constructed incorporating features of its cylindrical structure with nanopores in its walls. This model is used to study how ionic conductance along the lumen is affected by flux through the nanopores when an external potential is applied across its two ends. Based on the results of Brownian dynamics simulations, the nanopores were found to have asymmetric inner and outer conductances, manifested as nonlinear IV curves. Our simulations indicate that a combination of this asymmetry and an internal voltage source arising from the motion of the C-terminal tails causes a net current to be pumped across the microtubule wall and propagate down the microtubule through the lumen. This effect is demonstrated to enhance and add directly to the longitudinal current through the lumen resulting from an external voltage source, and could be significant in amplifying low-intensity endogenous currents within the cellular environment or as a nano-bioelectronic device.

**Introduction**

It has been proposed that cytoskeletal elements including actin and microtubules are involved in facilitating the propagation of electrical signals in the cell (see these review papers for an in-depth discussion: (1, 2)).

Priel *et al.* (3) have experimentally demonstrated that microtubules are excellent conductors of electrical signals and can amplify an electrical current stimulated by a pulse of applied voltage. The current collected 20-50 μm further down the microtubule (MT) was found to be double its value in the buffer solution with no MT present. Likewise Minoura and Muto (4) found using electroorientation experiments, that the conductivity of the MT was $1.5 \times 10^2$ mS/m, which was 15 times higher than that of the surrounding buffer. Similar results have been obtained for electrical signal conduction along actin filaments (5). Filamentous actin and MTs both share the characteristics of being protein biopolymers with highly negative linear charge densities, and thus both structures are surrounded by condensed atmospheres of counterions lining their outer surfaces, which would lead to high conductances along these ionic atmospheres. However, the structure of the MT is somewhat more complex, being cylindrical in shape with a hollow interior and an additional negative linear charge density along its inner surface and having a lattice-like wall interspersed with nanopores connecting the inner and outer surfaces. The question arises whether this cylindrical structure, and particularly its decorating nanopores, might be associated with novel bio-electronic properties that would contribute to the MT's proposed role in enhancing electrical signaling.

MTs self-assemble by polymerization of α/β-tubulin protein heterodimers (see Figure S1a). These building blocks are arranged linearly into protofilaments, and a ring of 13 protofilaments composes the wall of the cylindrical MT, which is typically several micrometers in length (and up to millimeters under special circumstances) and is 11.4 nm in radius at the midpoint of its wall. Each dimer in the MT has a length of 8.1 nm, a width of about 6.5 nm and a radial dimension of about 4.6 nm (6). A wide space, approximately 15 nm in diameter, known as the lumen, makes up the inner core of the cylinder.

The wall of a MT is generally characterized by what is known as a B-lattice structure, displaying 2 distinct types of nanopores (although it should be noted that there are 2 additional types of nanopores seen in the A-lattice structure). One nanopore in the B-lattice (nanopore 1) is located in the region where the interdimer $\beta/\alpha$ interface of one tubulin molecule lies next to the interdimer $\beta/\alpha$ interface of the adjacent tubulin molecule. A second nanopore (called nanopore 2) arises where the $\alpha/\beta$ intradimer interface of one tubulin molecule lies next to the $\alpha/\beta$ intradimer interface of the adjacent tubulin molecule (see Figure S1a-b). The effective radius at the narrowest point of the pore is approximately 4.0 Å for the type 1 pore, and 4.7 Å for the type 2 pore, when calculated by the HOLE program (7) with AMBER radii (8).

Little is known about the biological relevance of the nanopores or of the lumen, which are difficult to probe experimentally. Interestingly, chemotherapy drugs of the taxane family are believed to diffuse through the MT wall via the nanopores to a binding site known to be located in the lumen (9, 10). There is also evidence for binding of ligands within the nanopore; for example the MT-disrupting agent cyclostreptin has been shown to bind at a site within the MT nanopore (11). Additionally, the microtubule associated protein (MAP) tau has been observed binding to a site in the lumen (12, 13). Based on their size, the nanopores are expected to be quite permeable to the atmosphere of counterions covering both sides of the MT. The flexible carboxy-terminal tail domains of tubulin, consisting of approximately the last 10 residues on α-tubulin and 18 residues on β-tubulin, depending on the tubulin isotype, extend from the outer surface of the MT over the nanopores (see Figure S1a). These tail-domains have a high proportion of negatively charged residues; for example, the tail domain of βI tubulin has 9 Glu residues and 2 Asp residues, and that of αIV tubulin includes 5 Glu residues and 2 Asp residues, although the exact numbers vary slightly with isotype. This requires large numbers of counterions for charge compensation, and the fluctuations of the tail domains must be associated with accompanying movements of these counterions.

To the best of our knowledge no other study has yet focused on the motion of the surrounding counterions through the nanopores and around the MT, although this effect is a distinct possibility even just due to its geometrical similarity with ion flows across ion channels in membranes. Here, we intend to examine the flux of these cations under both equilibrium conditions, and under a constant perturbative electrostatic potential difference between the two ends of the MT. That is, we have constructed a physical model to calculate electric current through the sides of the MT, in addition to that flowing in the longitudinal direction along a MT. To parameterize this model, we have performed detailed atomic-scale calculations of the conductances of ions through a MT's nanopores using 3d Brownian dynamics. The results of our simulations will be presented in this paper and will hopefully guide future experimental validation of the predicted effect.

There is evidence in many cellular systems of an important and complex role for MTs in regulating ion-channel gating in the plasma membrane, although the mechanism of this influence has not been explained. For example, an intact MT structure has been shown to be essential for the activity of $GABA_A$ receptors in hippocampal neurons (14), for the functioning of polycystin-2 calcium channels in the kidney (15, 16), and has been associated with a decreased activation threshold of Na+ currents (17) and increased currents of $Na^+$ (18) and of $Ca^{2+}$ (19, 20) in neurons. On the other hand, in other studies, MT depolymerization has been associated with increased calcium currents in cardiac myocytes and in neurons (21-24). Studies also show that ion channels can be anchored to MTs through other proteins (14, 25). Thus electrically-excitable cells may possess ion channels which, by a somewhat different mechanism, have a similarly important dependence on MTs to that of mechano-sensitive channels, which are tethered by other proteins to MT bundles thought to regulate their opening and closing (26, 27).

The observations that MTs can regulate ion-channel activity, together with their remarkable conductive properties, and also the fact that both dendrites and axons in neurons are dense in MTs and moreover display distinctly-developed cytoskeletal architectures (28), suggest that MTs may play a role in electrical signal conduction in neurons. Various instances suggest that local amplification of ion concentration could be important during the extremely complex process of action potential generation. These include the generation of an axonal action potential by distributed processing of the synaptic activity at distinct dendritic sites (29); the observation of non-inactivating behavior in a small percentage of dendritic sodium channels (30); and the reduced voltage firing-threshold for channels located in the axonal hillock (31).

While there is as yet no experimental confirmation that this includes electrical signal conduction, an abundance of experimental evidence supports some role of MTs during neuronal signaling. In most tissues MTs are dynamic, growing and shrinking, but are less so in the neurons, where however rearrangement of MTs occurs during learning. Impaired learning has been found to correlate with decreased MAP2 levels (32-34), an excess of MAP tau (35, 36), and can be brought on by colchicine-induced changes in MT dynamics (37-39). Additionally, memory consolidation is accompanied by increased binding of tubulin to other neuronal proteins (40), by alterations in MAP2 expression (40-45), and by MAP 1B phosphorylation (46).

The remainder of this paper is divided into 3 sections. In the first, the various elements of the model we develop for ion propagation along MTs are described. Methods used to estimate the necessary parameters are described, including calculations of the electrical conductances of the two major types of pores in the MT wall by Brownian dynamics, and voltage differences arising from the fluctuating C-terminal domains. In the next section we present the parameter set derived for the circuit model, and give results. Finally, there is a conclusion section, in which results of our modeling are used to suggest future experimental and computational studies.

**Model development**

We have developed an electrical circuit model for the conduction of cations along a MT, with various electrical elements, which we describe below. Since anions are scarce in the intracellular environment and in the medium surrounding the negatively-charged MT in particular, we make the approximation that potassium cations are the only charge carriers present. The approximation is also made that the MT has cylindrical symmetry, and the volume surrounding the MT is divided into four concentric cylindrical shells, depicted as linear wires in Figure 1. Currents and voltages are cylindrically averaged over shells extending the length of one tubulin monomer along the MT. Because counterion clouds result in reduced resistivities, cylindrical shells are chosen to represent the condensed ionic atmospheres adjacent to the inner and outer surfaces, while outer bulk solution and the luminal volume are each modeled as separate electrical transmission lines. A set of resistances also regulates the numbers of cations allowed to cross between adjacent cylindrical shells in response to a voltage difference across the boundary between the two volumes. In Figure 1, the row of tubulin monomers, which has been sketched in, represents the wall of the MT.

Considering that the Bjerrum length, defined as the separation at which the electrostatic interaction energy equals thermal energy, is 7.13Å at a temperature of 293K (47), we chose to consider a 10Å-wide layer near the inner surface and a layer near the outer surface 40Å-wide, which is approximately 10Å extending beyond the positions of the C-termini.

*Conductances:* To determine conductances we made use of the well-known relation, based on the equivalent of Ohm's law for ionic flows

$$G = \frac{\sigma(c)A}{l} = \frac{\Lambda_0^{K^+} cA}{l}, \qquad (1)$$

where $c$ is the cationic concentration, $l$ is the distance in the direction of ionic flow, $\sigma(c)$ is the conductivity at a concentration $c$ of cations, $A$ is the cross-sectional area, and $\Lambda_0$ is the proportionality coefficient determining conductivity as a function of ionic concentration, which is approximately equal to 7.5 $(\Omega m)^{-1}M^{-1}$ for $K^+$ (48, 49). Further details of the evaluation of conductances are given in the supplemental section.

*Brownian dynamics:* The conductances of the nanopores occurring in the MT lattice limit the flow of ions between the inside and the outside of the MT. We made use of a Grand Canonical Monte Carlo program (GCMC/BD) created by the Roux group for computing channel conductances by Brownian dynamics simulations (50-53). This method has been shown to be successful at determining accurate current-voltage (I-V) curves for ion channels, especially those with sufficiently wide cross-sectional areas (> ~20 Å$^2$) to allow ion motion to be suitably described by Langevin dynamics (53, 54). The nanopores in the MT wall are comparable in area to ion channels to which the program has been successfully applied.

The program uses the Ermak-McCammon equation (55) to generate the dynamics of ions undergoing random solvent collisions. All interactions in the system are being described via a multi-ion potential of mean force (PMF) and thus the equations of motion can be expressed as:

$$\dot{\mathbf{r}}_i(t) = -\frac{D_i}{k_B T}\nabla_i W(\mathbf{r}_1,\mathbf{r}_2,\ldots) + \zeta_i(t), \qquad (2)$$

where $r_i$ is the position and $D_i$ is the diffusion constant of ion $i$, and $\zeta_i(t)$ is the Gaussian random noise. The multi-ion PMF, $W$, is given by (51, 52)

$$W(\mathbf{r}_1,\mathbf{r}_2,\ldots) = \sum_{i,j} u_{ij}\left(|\mathbf{r}_i - \mathbf{r}_j|\right) + \sum_i U_{core}(\mathbf{r}_i) + \Delta W_{sf}(\mathbf{r}_1,\mathbf{r}_2,\ldots), \qquad (3)$$

where $u_{ij}$ is the direct ion-ion interaction, $U_{core}$ is the repulsive potential arising from the excluded volume of the protein, and $\Delta W_{sf}$ is the electrostatic potential from the static charge distribution of the protein together with the electric field term $V$ (56, 57). $\Delta W_{sf}$ is determined as,

$$\Delta W_{sf}(\mathbf{r}_1, \mathbf{r}_2, \ldots) = \sum_i q_i \phi_{sf}(\mathbf{r}_i), \tag{4}$$

where the electrostatic potential $\phi_{sf}$ is calculated using a modified version of the Poisson-Boltzmann equation, known as the PB-V equation,

$$\nabla \cdot [\varepsilon(\mathbf{r})\nabla\phi(\mathbf{r})] - \bar{\kappa}^2(r)[\phi(\mathbf{r}) - V\Theta(\mathbf{r})] = -4\pi\lambda\rho^{prot}(\mathbf{r}). \tag{5}$$

Here $\Theta$ is the Heavyside step function equal to 0 for negative values of the z-coordinate of the position vector $\mathbf{r} = (x, y, z)$, and 1 for nonnegative z, where the z-axis is oriented through the pore and perpendicular to the axis of the MT; $\varepsilon$ is the space-dependent dielectric constant, set equal to 1 in the inner and buffer regions; $\bar{\kappa}^2$ is the space-dependent ionic screening factor, set equal to 0 in the inner and buffer regions; and $\rho^{prot}$ is the charge density of the protein (56, 57). Details of Brownian dynamics calculations are given in the supplemental section.

Some explanation is in order concerning the source of the electric field term $V$. In applications to which GCMC/BD has been applied in the literature (50-54), the voltage $V$ represents a membrane potential arising from a difference in local ionic composition on either side of the membrane in which the channel is embedded, and is accounted for by equation (5). Here, $V$ is accounted for by the same formalism, but represents, rather, a potential gradient across the nanopore in the MT wall arising from the fluctuating positions of the C-terminal tail domains and the fact that the number of ions screening the charges on these tubulin domains can be either in excess or insufficient. In what follows, this issue will be addressed in greater detail.

*C-terminal tails*: The charge of the C-terminal β-tubulin tail is generally -12 or -11 elementary charges, depending upon the β-tubulin isotype, and that of the α-tubulin (isotype IV) tail is -7 charges. Although the remainder of the tubulin dimer is also negatively charged (with a total of ~50 negative charges), each tubulin monomer has a positively charged cluster of residues, which may interact with the C-terminus of the neighboring monomer (58).

We have performed an investigation of potential tail-body interactions of the C-terminal tails with positively charged residues on the surface of the tubulin dimer by molecular dynamics simulation. Briefly, we first employed a molecular dynamics simulated annealing approach to determine tail conformations. Then, 15 high temperature implicitly-solvated MT dimers were generated at a temperature of 5,000K and were then slowly cooled to a target temperature of 300K. Our annealing schedule used steps of 1,000K over periods of 400ps until the temperature was below 2000K, and then used a 0.8-power law cooling-schedule. The tubulin dimers were then placed into a MT lattice, and following equilibration, molecular dynamics simulations were performed over periods of 10 ns, with periodic boundary conditions at a temperature of 300K. The details of these simulations will be published separately. However, it was found that C-terminal tails can exist in multiple states involving different numbers of salt bridges, and that the tails dynamically oscillate between these states as tail-body interactions form and then give way to interactions of tail residues with water and ions, with several such switches occurring within a single nanosecond.

When the negatively charged residues of the tails interact with positively charged residues on the tubulin dimer, fewer cations in the surrounding solvent are required to stabilize the negative charges on the protein, by comparison to when the tail residues are more extended away from the MT surface. There is also expected to be a corresponding decrease in the equilibrium number of surrounding anions, which however are not considered in our model. Since the exact equilibrium number of counterions within the condensed ionic atmosphere surrounding the outer surface of the MT varies as the positions of the C-terminal tails undergo thermal fluctuations, the chemical potential oscillates stochastically between being attractive or repulsive for the influx of additional cations. Thus the motion of the C-terminal tails as they alternately attach and detach from the MT surface may be viewed as a stochastic voltage source acting on the surrounding cations.

To estimate the voltage difference corresponding to the variation in the C-terminal tail position, we performed calculations of the capacitance of the volume corresponding to the outer counterion sheath. This capacitance was defined as the coefficient of proportionality between the number of $K^+$ ions in this outer ionic atmosphere and a potential placed between this layer and the outer bulk solution. Ion counts for a range of potential differences were calculated by solving the PB-V equation (5) as implemented in the PB-PNP program (51, 52). Further computational details are given in the supplemental section. The resulting curves for each pore type shown in Figure S2a and b are highly linear, and the two have a nearly identical slope corresponding to a capacitance of $1.15 \times 10^{-8}$ nF.

With a variation of the tail-body interactions, corresponding to different positions of the C-terminal tails, the linear plot of equilibrium ion-count versus potential is expected to shift up or down. If a single additional cation beyond the equilibrium number is present within the outer counterion sheath after a C-terminal tail has switched position to interact with the tubulin body, the line is shifted down by 1$e$. This corresponds to a repulsive voltage equal to 1$e$ divided by the capacitance, or ~14mV. Likewise when the C-terminal tail switches to a more extended conformation, there is an attractive potential of the same magnitude. In reality, a continuum of many different states is taken on by the C-terminal tails. Thus the thermal fluctuations of the C-terminal tails on a ring of 13 monomers in the MT can be modeled as a stochastically alternating battery with a maximum voltage magnitude of $\Delta n * 7$ mV, where $\Delta n$ is the variation in the equilibrium number of counterions in the condensed ionic atmosphere surrounding the ring. The frequency of oscillation was chosen to be 10/ns.

    Note that the assertion that a current is driven by the stochastic motion of the C-terminal tails in no way represents a violation of energy conservation, as molecular dynamics simulations clearly demonstrate that thermal energy is sufficient to drive large-scale fluctuations in the positions of these domains.

*Electrical circuit:* A circuit model was created to represent a MT with 20 monomers along its length using the Simulink and Simpowersystems modules of Matlab R2009a (59). Resistances, as well as both the external voltage and the stochastic alternating voltage and its frequency, were read as inputs. The program solved for voltages and currents at each segment of the circuit and propagated these in time, using a time step of 1ns. A switch, or circuit breaker, seen in Figure 1, was implemented to connect the circuit to the external voltage source, and opens at time zero to allow current to cross. The sources of time dependence in the model consisted of a stochastic battery representing the motion of the C-terminal tails as described above, as well as the integrators used to determine net charge transfer. The random battery generates uniformly distributed numbers over an interval between the positive and negative values of its specified voltage, with a specifiable starting seed. Because the time-dependent current showed large stochastic oscillations, its time-integration was carried out to determine the amount of net charge transferred as a function of time at the end of the MT held at lower potential, in each of the 4 cylindrical regions.

**Results**

*Pore conductances determined by Brownian dynamics:* Calculated currents as a function of voltage difference for the conduction of cations across the nanopore of type 1 are plotted in Figure 2. The IV curve showing the conduction of anions through this pore, as well as curves depicting the conduction of positive and negative ions through the nanopore of type 2 is shown in Figure S3. Error bars representing standard deviations, also shown in the figure, were at most 12 pA, 9.0 pA, 32 pA, and 3.7 pA for Figure 2, and Figure S3 parts (a), (b), and (c), respectively. Conductances through nanopores were determined from the slopes of linear fits of the IV curves shown in the figures. Both types of pores show asymmetric inner and outer conductances of cations. For the type 1 pore, inner and outer cationic conductances were found to be 2.93 nS and 1.22 nS, respectively. The conductance of cations was found to be somewhat greater for the type 2 pore, with calculated values of 7.80 nS and 4.98 nS, respectively, for inner and outer conductances. The cylindrical shell-like wall of the MT has a sandwich-like electrostatic potential being negative on the inside, even more negative on the outside, and positive in the center (see

Figure S1b). The very negative potential on the outer surface attracts cations into the pore at a positive potential, and also makes it difficult for anions to enter the pore at a negative potential.

Both nanopores were found to be almost completely impermeable to anions at negative potentials (43.7 pS and 11.8 pS for type 1 and 2 pores, respectively). The positive parts of the anion conductance curves in Figures S3a and c are fit well by exponential curves, as shown. Conductances of anions into the pores increase substantially as the voltage is raised above 150 mV, from about 11.8 pS to 198 pS for the type 2 pore, and from 240pS to 1.10 nS for the type 2 pore, as determined from linear fits of the positive voltage data in the intervals between 0 and 150 mV, and between 150mV and 250mV, respectively. The difficulty for anions to enter the pore is most likely partially due to the composition of the constriction zone. As determined by the HOLE program (7), the constriction zone of pore 1 is bounded by α-tubulin residue THR 130, by PHE 212 of the adjacent β tubulin, and ASP 327 of the other α tubulin. In pore 2, the constriction zone is bounded by residues THR 94 of α tubulin, ASP 128 of the adjacent β tubulin, and ASP 218 of the other α tubulin. Thus, the negatively charged ASP residues may make it difficult for anions to breach this area, and especially so in the type pore 2 where there are two such charged residues.

Asymmetric conductances have been measured in ion channels by GC/MCBD in the past, giving good agreement with experiment (50). For example, for the OmpF pore, conductances were calculated to be 1.36 nS and 1.15 nS with positive and negative voltages across the channel, respectively, compared to values of 1.25nS and 1.13 nS determined experimentally (53).

To assess the role of the protein charge distribution in contributing to nanopore conductances, calculations were repeated for the type 2 pore, but this time with all protein charges turned off. We found that this led to the cation conductance being reduced by a factor of about 15. The conductance of anions was much larger than for the charged model, being increased by a factor of about 24 at a negative potential, and by a factor of about 4 at high positive potentials. These results show that electrostatics has a very significant effect on the conductive properties of the nanopores. We also investigated the effect of a different C-terminal tail conformation; with a different position with respect to the MT, the conductance of cations was changed by only 11%, and there was very little effect on the conductance of anions.

Within our circuit model, a pair of forward-biased diodes was used to model the asymmetric conductance of each MT pore, with one controlling the flow of cations into the MT, and the other regulating the outward flux, as is illustrated in Figure 1.

*Parameterization of electrical circuit:* The repeating unit of 2 monomers is illustrated in Figure 1. Values of resistances, which were calculated as described above for use in our model, are given in Table I. The bulk resistance shown in the table was calculated using an outer radius of 50 nm for the entire system. This value was chosen based on the typical distance of 100 nm separating parallel MTs in the cell such as a neuron. Resistance is smallest in the bulk region, followed by the outer condensed ionic atmosphere, and the inner condensed ionic atmosphere, with the inner lumen having the smallest conductance.

*Application of circuit model to dynamics of ionic flow:* The integrated currents running parallel to the MT axis in each of the four regions, are plotted in Figure 3. A value of 28 mV was used for the stochastic voltage source and 1mV, for the external voltage source applied across the two ends of the MT.

<u>Asymmetric conduction of microtubule pores enhances ionic flow through the lumen.</u> The elements of the circuit model that confer nonlinearity are the asymmetric conductances of the nanopores and the stochastic tail battery. To determine what effect these have on the longitudinal currents, we compared results for the case where there is no internal voltage from the tail movement (see Figure S4), when there is no asymmetry in the conductance in Figure S5, and when there is neither asymmetry nor an internal voltage. The same values corresponding to inward resistances were assigned to the outer resistances to model symmetric conduction.

Without the voltage created by the stochastic motion of the C-terminal tails, plots of charge transferred versus time are linear, indicating a constant current. In this case it makes no difference whether the conductance of the nanopores is symmetric (not shown) or asymmetric, as there is no driving force on the ions in the lateral direction. The net charge transferred in the bulk region after the end of 1000 ns is $1.0 \times 10^{-7}$ nC, while a net charge of about $1.1 \times 10^{-8}$ nC flows along the outer ionic atmosphere, which has a smaller area although its conductivity is greater. Only $4 \times 10^{-9}$ nC is transferred along the inner ionic atmosphere of the MT and even less ($3 \times 10^{-9}$ nC) through the central lumen.

When the C-terminal tail fluctuation is accounted for in the model but with symmetric pore conduction, the summed charge transferred along the inner ionic atmosphere and the lumen is $3.1 \times 10^{-10}$ nC at the end of the 1000 ns time period. Small fluctuations on the time-scale of the tail motions are exhibited in the plots, as well as larger time-scale fluctuations, representing an oscillatory exchange of ions between the bulk solution and the condensed atmosphere of counterions near the MT´s surface.

When the conductance asymmetry of the pores is included in the model, the total charge transfer in the lumen increases to about $4.4 \times 10^{-8}$ nC, with $2.6 \times 10^{-8}$ nC moving along the inner ionic atmosphere and $1.8 \times 10^{-8}$ nC moving through the inner lumen. It may be noted by comparing values at 500 ns to those at 1000 ns that the average currents in the two luminal regions are well-converged by 1000 ns. Thus the overall charge transferred divided by the total time may be thought of as the average current, and the total average current through the lumen predicted by our circuit model is $4.4 \times 10^{-2}$ nA for Figure 3. These results demonstrate that the conductance asymmetry of the MT wall allows the random internal battery to pump cations transversely through the nanopores into the lumen, from which they are directed out of one of the open ends of the MT.

Total longitudinal charge transferred along MT is independent of nanopores. In each of Figures 3 and S4-S5, the value of total charge transferred after 1000 ns is identical. Therefore, the summed current along the outer ionic atmosphere and in the bulk, which without the asymmetric conductance or the internal random battery had been $1.1 \times 10^{-1}$ nA, is reduced by the additional $3.7 \times 10^{-2}$ nA current which is now shunted through the lumen; that is, the additional luminal current is drawn from the outer ionic atmosphere and the outer bulk solution.

2D spatial-temporal plots reveal that net ionic current flows inward through some pores and outward through others. Net charge transfer was determined as a function of position along the MT by integrating current outputs after each of the 20 tubulin monomers, to yield 2D spatial-temporal plots. A random battery voltage of 28 mV was used, and external voltages of 0mV and 1mV, as shown in Figure 4 and Figure S6, respectively. The average currents at the end of the 1000 steps, given by the overall charge transferred divided by total time, are directed to the right where the area is colored red or yellow and to the left if the coloring is blue. An interesting pattern of currents arises surrounding the MT's wall, in the 0mV case. Close to the ends of the MT the current flows into the MT and out through the open ends. On either side of the center, the current follows a loop, so that there are currents flowing away from the center on the outside of the MT and towards the center on the inside. The pattern of currents is symmetric about the center of the MT. The current driven through the wall of the MT leaves either one or the other end of the MT with equal probability. With the 1mV applied voltage, within the lumen there is a multidirectional spatial gradient of charge transfer, whereas in the bulk solution at any given time the direction of charge flow is much more uniform along the length of the MT.

<u>Net current driven through pores adds directly to that induced by the external potential, independent of value of external potential or MT length.</u> In Figure 5, the current-voltage (IV) curves are shown for total luminal current versus external voltage, for values of the random battery equal to 14 mV and 28 mV, respectively. Simulations were run for values of the external voltage ranging from -100 to 100 mV. Plots were determined to be linear over the entire range of voltages. In the case of symmetric conductance, the IV curve passes through the origin. Asymmetric conductance of the pores shifts the curve towards positive currents. However, the slope is unchanged, showing that the current being driven through the nanopores adds directly to the current induced by the external potential. The effect of varying the length of the MT was also examined, using MTs composed of 10 or 30 monomers (see Figure S7). The y-intercepts of the resulting plots both coincide with that obtained for a length of 20 monomers, although the slope changes as a result of the altered voltage per unit distance. That is, additional tail batteries associated with a longer MT cannot increase the transverse current across the MT wall and the same net current is pumped across independent of length.

Net current through pores is roughly linear in the value of the tail battery. The positive shifts of the IV curves given by the y-intercepts of the plots in Figure 5 were 0.021nA and 0.037nA for random battery voltages of 14 mV and 28 mV, respectively. If the random battery is increased to 140 mV, then the shift is 0.19 mV. Thus the value of the current scales almost linearly with the unknown random voltage.

Net luminal current driven by fluctuations of C-terminal tails exceeds that driven by the external battery at small values of the external voltage. At zero or even small negative values of the voltage, a positive current persists at the end of the MT, just as there was a backwards flow of charges at the opposite end for small positive values of the voltage. The x-intercept on each plot in Figure 5 or Figure S7 corresponds to, not only the external voltage at which forward current vanishes, but also the magnitude of external voltage required to halt the backwards flow of ionic current. Most importantly, for positive external voltages smaller than the absolute value of the x-intercept on the plot, the potential associated with the stochastic fluctuation of C-terminal tails is the dominant driving force acting on the cations, whereas as the external potential is increased beyond this point the stochastic tail voltage, while still adding to the total longitudinal current, contributes less than the external battery to this total. In summary, the addition to the longitudinal current by the current through the MT wall is most important at small values of the external voltage.

*Theoretical justification:* Our computational model demonstrates that the influx of cations through the nanopores results in an addition to the net transfer of cations to the more negative end of the MT. We now attempt to provide some basic understanding of this effect by using a simple theoretical model.

Consider the *n*th unit of a MT, shown in Figure 6. Here $I_n^b$, $I_n^{os}$, $I_n^{is}$ and $I_n^l$ are the current in the bulk (b), outer surface (os), inner surface (is) and the lumen (l), respectively. The random tail batteries $\varepsilon_n(t)$ and $\varepsilon_{n+1}(t)$ are time dependent and could be in different phases, and so in general $\varepsilon_n(t)$ will not equal $\varepsilon_{n+1}(t)$ in both magnitude and sign. The time dependent resistor blocks $R_n(t)$ and $R_{n+1}(t)$ are determined by the two sets of resistances $R_{p1}^{in}$ and $R_{p1}^{out}$, and $R_{p2}^{in}$ and $R_{p2}^{out}$, respectively:

$$R_n(t) = (1/2) [ R_{p2}^{in} (1+ \varepsilon_n/|\varepsilon_n|) + R_{p2}^{out} (1- \varepsilon_n/|\varepsilon_n|) ], \qquad (6)$$

$$R_{n+1}(t) = (1/2) [ R_{p1}^{in} (1+ \varepsilon_{n+1}/|\varepsilon_{n+1}|) + R_{p1}^{out} (1- \varepsilon_{n+1}/|\varepsilon_{n+1}|) ]. \qquad (7)$$

The direction of the current in the HG (CD) branch depends on whether the tail battery $\varepsilon_n$ ($\varepsilon_{n+1}$) is pushing the charge carriers through the nanopores, or pulling out from them. As a result, the current flows in or out through different resistors. The term $\varepsilon_n / |\varepsilon_n|$ in the above equation represents the dependence of these resistances on the state of the tail battery. In the symmetric case $R_{p2}^{out} = R_{p2}^{in} = R_{p1}^{out} = R_{p1}^{in}$, and so $R_n(t) = R_{n+1}(t)$.

To extract some information on the current or the resulting accumulated charges in the lumen, Kirchhoff's rules are applied:

Loop ABCHA:

$$\varepsilon_n(t) - \varepsilon_{n+1}(t) - (R_b + 2 R_{outer}) I_n^b + R_{outer} (I_{n-1}^b + I_{n+1}^b) + R_{os} I_n^{os} = 0, \qquad (8a)$$

Loop HCDGH:

$$R_{is} I_n^{is} - (R_{os} + R_n + R_{n+1}) I_n^{os} + R_n I_{n-1}^{os} + R_{n+1} I_{n+1}^{os} = 0, \qquad (8b)$$

Loop GDEFG:

$$R_{lumen} I_n^l - (R_{is} + 2 R_{inner}) I_n^{is} + R_{inner} (I_{n-1}^{is} + I_{n+1}^{is}) = 0. \qquad (8c)$$

The difference $\Delta I_n^l = I_{n+1}^l - I_n^l$ represents the change in the current through the lumen from the *n*th unit to (*n+1*)th unit, which can be shown by combining Eqs. (8a) to (8c) to satisfy,

$$R_{os} R_{lumen} \Delta I_n^l = [R_{os} - \Delta^2 R_n(t)] (\Delta^2 \varepsilon_n(t) + R_b \Delta I_n^b), \qquad (9)$$

where $\Delta F_n = F_{n+1} - F_n$ and $\Delta^2 F_n = F_{n+1} - 2 F_n + F_{n-1}$ for any function $F_n$. For small values of the external battery one expects that the bulk current does not change significantly and so $\Delta I_n^b \sim 0$. Therefore,

$$R_{os} R_{lumen} \Delta I_n^l \sim [R_{os} - \Delta^2 R_n(t)] \Delta^2 \varepsilon_n(t), \qquad (10)$$

In the symmetric case $R_{n+2}(t) = R_{n+1}(t) = R_n(t)$ and so $\Delta^2 R_n(t) = 0$:

$$R_{lumen} \Delta I_n^l \sim \Delta^2 \varepsilon_n(t), \qquad (11)$$

or

$$R_{lumen} (d/dt) \Delta Q_n^l \sim \Delta^2 \varepsilon_n(t), \qquad (12)$$

where $\Delta Q_n^l$ is the change in the accumulated charges in the lumen between units n and n+1. Since the tail battery $\varepsilon_n(t)$ is random in nature, its average value over the course of time should be zero. Therefore, $\Delta Q_n^l \sim 0$ and so one would not expect to see a significant change in the average value of the current through the lumen or in the resulting accumulated electric charges.

In the asymmetric case, however, $\Delta^2 R_n(t)$ is not zero and so one has

$$R_{os} R_{lumen} \Delta Q_n^1 \sim - \int dt \, \Delta^2 R_n(t) \, \Delta^2 \varepsilon_n(t). \tag{13}$$

It is clear that the above integrand has zeroth and second order terms in $\varepsilon_n(t)$:

$$\Delta^2 R_n(t) \Delta^2 \varepsilon_n(t) = [R_{p2}^{in} + R_{p2}^{out} - R_{p1}^{in} - R_{p1}^{out}]$$

$$+ (1/2)(R_{p2}^{in} - R_{p2}^{out})(\varepsilon_{n+1}/|\varepsilon_{n+1}| + \varepsilon_{n-1}/|\varepsilon_{n-1}|) - (R_{p1}^{in} - R_{p1}^{out}) \varepsilon_n/|\varepsilon_n|]$$

$$\times (\varepsilon_{n+1} - 2\varepsilon_n + \varepsilon_{n-1}) \tag{14}$$

As a result, one expects the average accumulated charge to be nonzero over the course of time as we move along the MT units.

**Conclusions**

The model presented here demonstrates how the MT's distinctive geometry, being a cylindrical biopolymer with channel-like nanopores and C-terminal tails extending outward from the surface, may enhance the conduction of cations through the lumen, as a result of asymmetry in ion conduction through the pores and fluctuating tail-body interactions. However, our treatment of the difference in potential between the two ends of the MT as fixed entails that, within this model, the sum of longitudinal currents through all four cylindrical regions is necessarily constant, independent of the presence or absence of conduction through the sides of the MT. The type of situation that is more likely to arise in the cellular environment is a short-lived state of increased potential at one end of the MT, slowly dispersing as ions flow towards the other end. In that case it may be expected that transferred charges would accumulate at the end with the more negative potential, and that current passing through the lumen would not all back-propagate in the opposite direction along the bulk and outer surface, as in the results presented here. That is, the total longitudinal current may increase as a result of the current through the pores. In future work, a more realistic computational model allowing for dynamic equilibration, for example by solution of the Smoluchowski equation, could be used to test this hypothesis. From a biological perspective, this could imply that when a weak potential difference occurs across its two ends, a MT may allow the resulting current to be amplified along its direction, preventing dissipation as random noise throughout the cell. It is possible that within the cellular environment this may contribute to the opening of voltage-gated ion channels with which the MTs form a terminal connection.

It is also tantalizing to speculate about the role of the most structurally variable element of tubulin, its C-terminal tail region, that can be thought of as a tubulin isotype identifier. With major differences across the isotypes in both the length and charge of C-termini, their specialized role as an active dynamical element of the electrical circuitry formed by a MT may provide an explanation for tubulin isotype distribution differences among different cell types, e.g. the abundance of βIII tubulin in neurons (60).

Interestingly in this connection, Minoura and Muto (4) found that when the C-terminal tail region of the MT was cleaved by subtilisin digestion or when the pH was varied, the effect on conductance was proportional to the decrease in the net number of negative charges, indicating that the high conductivity was primarily due to the polarization response of the condensed atmosphere of counterions to the external voltage source. The volume for which conductance was calculated was 21.6 nm in radius, and thus would have included a region of what we have referred to as the outer bulk solution. These observations agree with our results predicting that the stochastic motion of C-terminal tails has no effect on overall charge conduction. However, we expect that the 'pipe-like' conduction of cations through the lumen predicted here would increase the current focused at a narrow region at one end of the MT. Experimental testing of the results of the present model would require a narrowly-focused current sensor capable of recording the isolated luminal current. Digestion of C-terminal tails by subtilisin or changes in pH or ionic concentrations, could then be used to test our predictions of the involvement of C-terminal tails by ascertaining how MT conductivity through the lumen varies with the corresponding net negative charge of the dimer. The much-needed experimental validation of our predictions would undoubtedly lead to exciting new opportunities in the development of biological nano-scale electronic applications.

**Acknowledgements.** We thank Dr. Tyler Luchko for help with preparing microtubule models, Dr. Ansgar Philippsen for assistance with visualizing BD trajectories, and Prof. Tim J. Lewis for discussions. This work was supported by grants to J.A.T. from the Allard Foundation, the Alberta Cancer Foundation, and NSERC; and by Pacific Institute of Mathematical Sciences and Bhatia foundation fellowships to HF. S.Y.N. is supported by a Discovery Grant RGPIN-315019 from the Natural Sciences and Engineering Research Council of Canada (NSERC). S.Y.N. is an Alberta Heritage Foundation for Medical Research (AHFMR) Scholar and Canadian Institute for Health Research New Investigator. This work was made possible by the use of the West-Grid computing facilities.

|                | inner lumen | inner sheath           | outer sheath         | outer bulk |
|----------------|-------------|------------------------|----------------------|------------|
| **inner lumen**    | $1.77 \times 10^7$ | $1.07 \times 10^6$ |                      |            |
| **inner sheath**   | $1.07 \times 10^6$ | $1.25 \times 10^7$ | $6.41 \times 10^7$ $1.54 \times 10^7$ |            |
| **outer sheath**   |             | $2.65 \times 10^7$ $9.86 \times 10^6$ | $4.75 \times 10^6$ | $4.06 \times 10^6$ |
| **outer bulk**     |             |                        | $4.06 \times 10^6$ | $5.31 \times 10^5$ |

**Table I.** Computed resistances in Ohms. Diagonal elements describe the flow of $K^+$ ions along cylindrical shells running parallel to the microtubule axis in each of 4 regions. Values corresponding to flow between adjacent volumes are shown as off-diagonal elements, with the region of positive potential given in the first column; upper and lower values represent flow through pore 1 and pore 2, respectively.

**Figure Captions.**

Figure 1: 2-dimensional electrical circuit model of the microtubule. The vertical direction is radial, and the horizontal direction is along the axis of the microtubule. The various electrical resistances, in addition to the voltage sources, are labeled to show the physical placements of these components relative to the microtubule in the model. The abbreviations "o.s." and "i.s." are used to denote the ionic atmospheres close to the outer surface and inner surface, respectively; "ax" stands for axial, and "rad" for radial.

Figure 2: Current-voltage relation from GCMC/BD simulation for conductance of cations through the type 1 pore. Error bars represent standard deviations over 5 independent data subsets.

Figure 3: Time-integrated current (nC) for asymmetric pore conductance plotted against time (ns), with an external voltage of 1 mV, and a stochastic voltage of 28 mV. The four panels, are from top to bottom, time-integrated current through (a) the bulk solution outside the microtubule, (b) the outer ionic atmosphere of the microtubule, (c) the inner ionic atmosphere of the microtubule, and (d) the rest of the lumen.

Figure 4: Color plot showing integrated current (nC) as a function of time (ns) and tubulin unit through (a) the bulk solution outside the microtubule, (b) the outer ionic atmosphere of the microtubule, (c) the inner ionic atmosphere of the microtubule, and (d)

the rest of the lumen.. A stochastic voltage of 28 mV was used, and an external voltage of 0 mV.

Figure 5: Current-voltage relations calculated using the circuit model for conductance of cations through the microtubule lumen. A microtubule length of 20 monomers was used. The dotted line corresponds to a stochastic voltage of 14mV, and the dashed line to a stochastic voltage of 28mV. The solid line corresponds to symmetric pore conductance.

Figure 6: The *n*th unit of the same microtubule model shown in Figure 1.

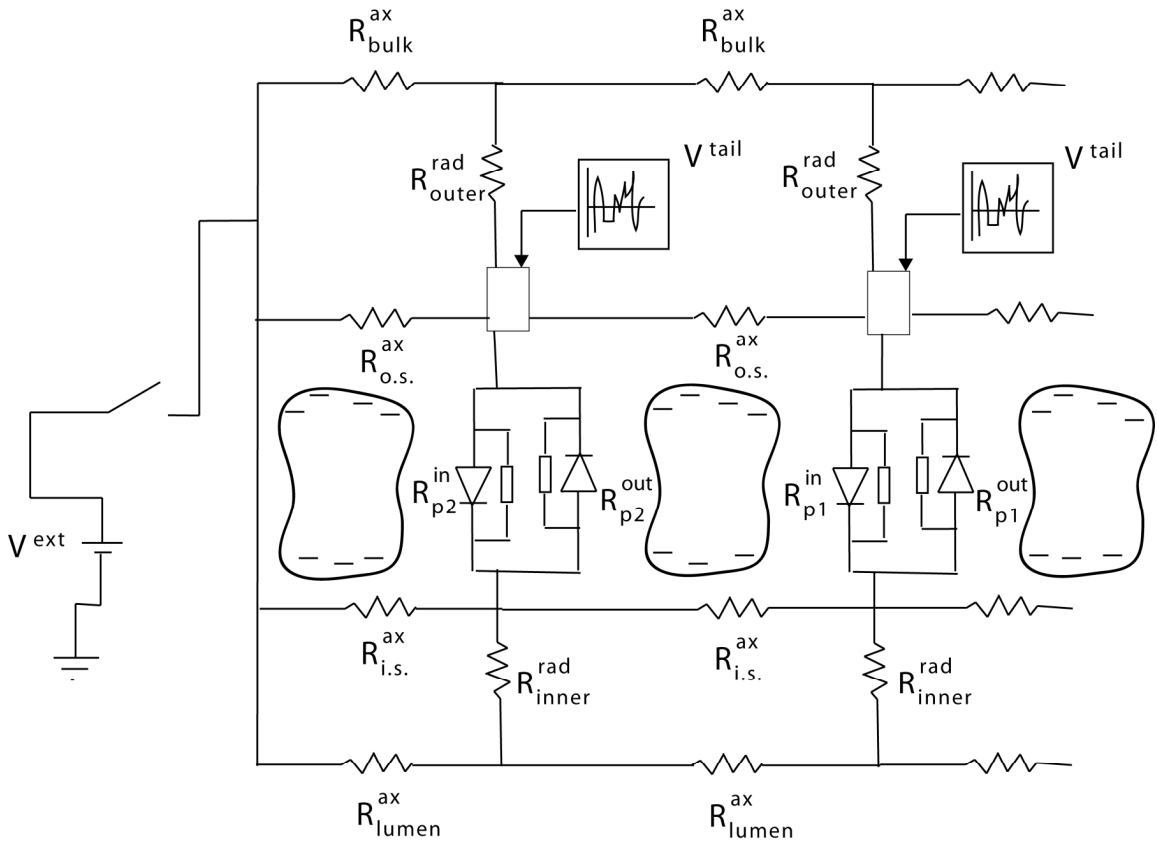

**Fig. 1**

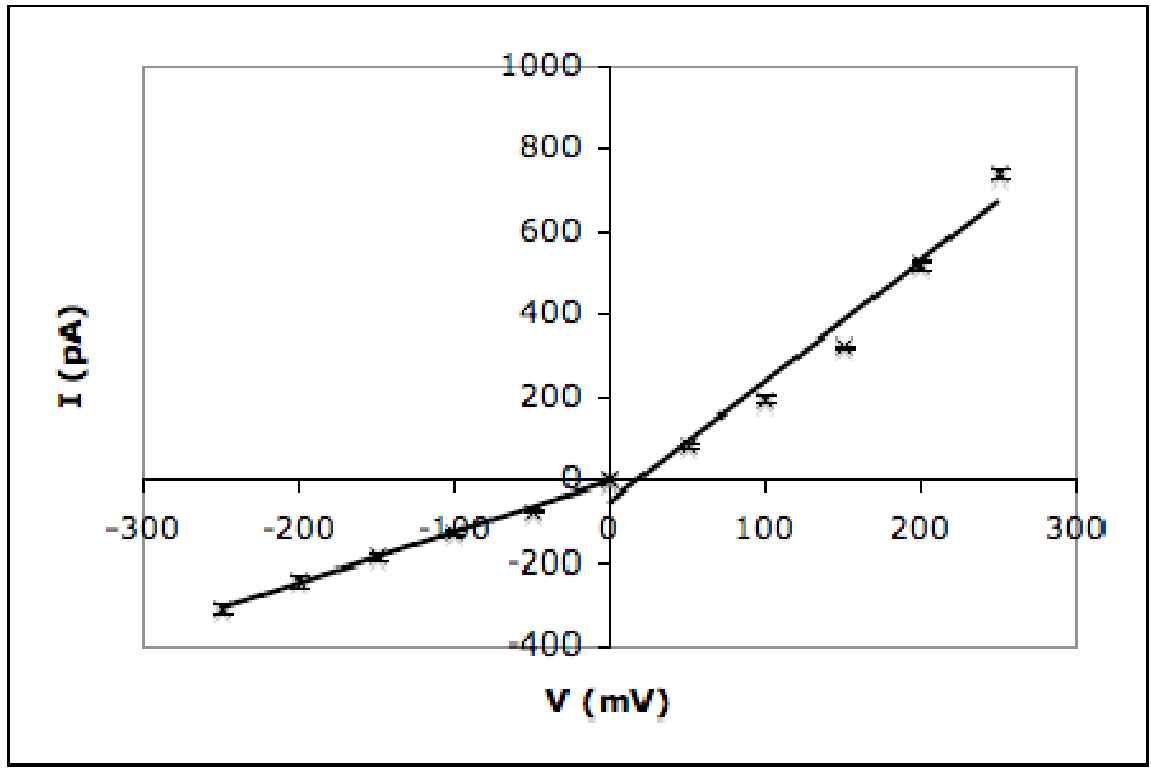

**Fig. 2**

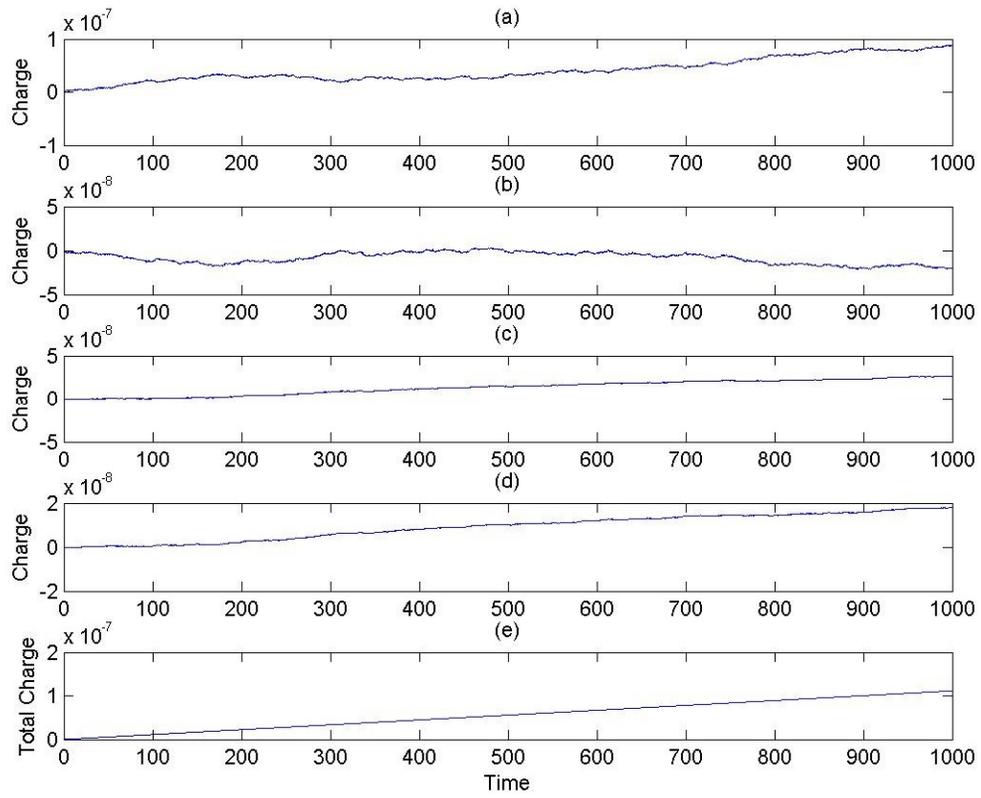

**Fig. 3**

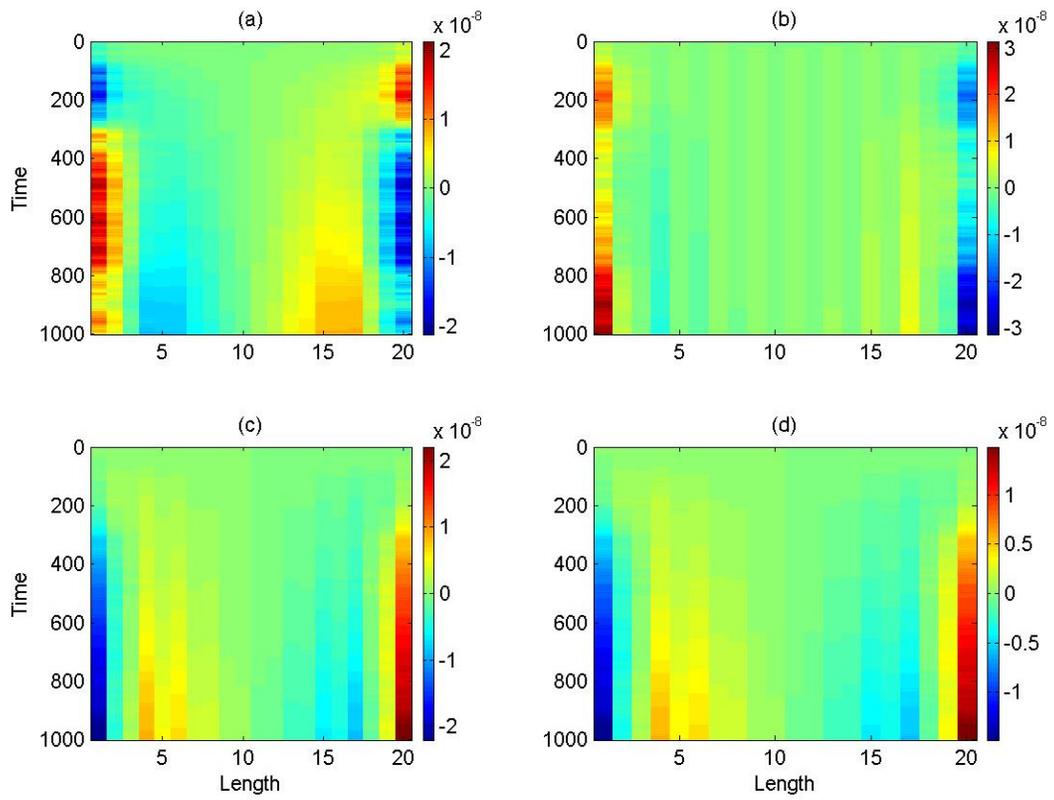

**Fig. 4**

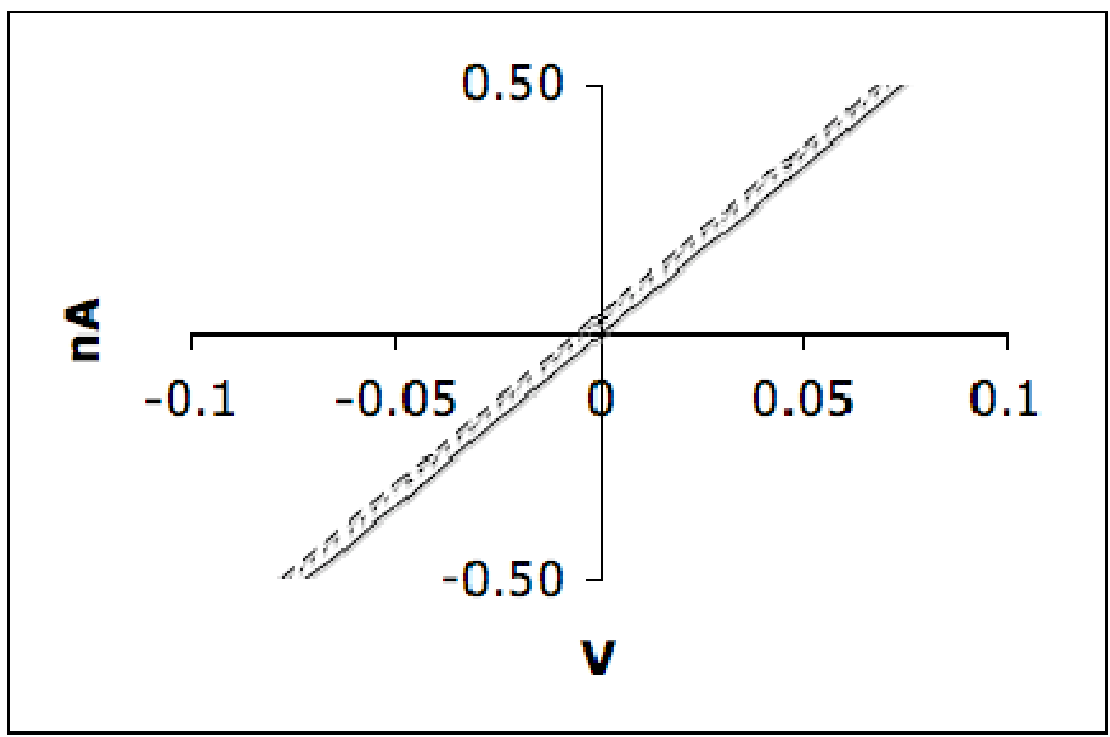

**Fig. 5**

**Fig 6**